\documentstyle[epsf,eqsecnum,floats,preprint,aps]{revtex}

\tighten

\input epsf
\begin{document}

\newcommand{\be}{\begin{equation}}
\newcommand{\ee}{\end{equation}}
\newcommand{\bea}{\begin{eqnarray}}
\newcommand{\eea}{\end{eqnarray}}
\newcommand{\PSbox}[3]{\mbox{\rule{0in}{#3}\includegraphics{#1}\hspace{#2}}}

\def\5M{M^3_{(5)}}
\def\4M{M^2_{(4)}}

\overfullrule=0pt
\def\Int{\int_{r_H}^\infty}
\def\d{\partial}
\def\e{\epsilon}
\def\M{{\cal M}}
\def\high{\vphantom{\Biggl(}\displaystyle}
\catcode`@=11
\def\@versim#1#2{\lower.7\p@\vbox{\baselineskip\z@skip\lineskip-.5\p@
    \ialign{$\m@th#1\hfil##\hfil$\crcr#2\crcr\sim\crcr}}}
\def\simge{\mathrel{\mathpalette\@versim>}} % 
\def\simle{\mathrel{\mathpalette\@versim<}} % 
\catcode`@=12 % at signs are no longer letters

\rightline{hep-th/0208169}
\vskip 4cm

\setcounter{footnote}{0}

\begin{center}
\large{\bf Global Structure of Deffayet
	(Dvali--Gabadadze--Porrati) Cosmologies}
\ \\
\ \\
\normalsize{Arthur Lue\footnote{E-mail:  lue@physics.nyu.edu}}
\ \\
\ \\
\small{\em Center for Cosmology and Particle Physics\\
Department of Physics\\
New York University \\
New York, NY 10003}

\end{center}

\begin{abstract}

\noindent
We detail the global structure of the five-dimensional bulk for the
cosmological evolution of Dvali-Gabadadze-Porrati braneworlds.  The
picture articulated here provides a framework and intuition for
understanding how metric perturbations leave (and possibly reenter)
the brane universe.  A bulk observer sees the braneworld as a
relativistically expanding bubble, viewed either from the interior (in
the case of the Friedmann--Lema\^{\i}tre--Robertson--Walker phase) or
the exterior (the self-accelerating phase).  Shortcuts through the
bulk in the first phase can lead to an apparent brane causality
violation and provide an opportunity for the evasion of the horizon
problem found in conventional four-dimensional cosmologies.  Features
of the global geometry in the latter phase anticipate a depletion of
power for linear metric perturbations on large scales.
\end{abstract}

\setcounter{page}{0}
\thispagestyle{empty}
\maketitle

\eject

\vfill

\baselineskip 18pt plus 2pt minus 2pt

\section{Introduction}

The gravity theory of Dvali--Gabadadze--Porrati (DGP) is a braneworld
theory with a metastable four-dimensional graviton
\cite{Dvali:2000hr}.  The graviton is pinned to a four-dimensional
braneworld by intrinsic curvature terms induced by quantum matter
fluctuations; but as it propagates over large distances, the graviton
eventually evaporates off the brane into an infinite volume,
five-dimensional Minkowski bulk.  As a result, the DGP braneworld
theory is a model in a class of theories in which gravity deviates
from conventional Einstein gravity not at short distances (as in more
familiar braneworld theories), but rather at long distances.  Such a
model has both intriguing phenomenological
\cite{Dvali:2001gm,Dvali:2001gx,Lue:2001gc,Gruzinov:2001hp} as well as
cosmological consequences
\cite{Deffayet,Deffayet:2001pu,Deffayet:2002sp,Alcaniz:2002qh,Jain:2002di,Alcaniz:2002qm}.

A braneworld model of the sort where gravity is modified at extremely
large scales is motivated by the desire to ascertain how our
understanding of cosmology may be refined by the presence of extra
dimensions.  Observational pillars of the standard cosmological model,
including the cosmic microwave background and large scale structure,
provide important tests of theories that seriously modify physics at
cosmological scales.  Deffayet's cosmological equations for the DGP
model already point to important deviations from the standard model
\cite{Deffayet:2001pu,Deffayet:2002sp,Alcaniz:2002qh,Jain:2002di,Alcaniz:2002qm}.
In order to constrain the DGP theory further, understanding the
development of large scale structure is necessary.  Leakage of
gravitational energy into the bulk is a key aspect of the DGP
braneworld model, and one expects such leakage to modify the spectrum
of density fluctuations at the largest observable cosmological scales.
However, a detailed analysis, or even an intuitive understanding, of
how spectral power of metric perturbations on the brane fills the bulk
and (potentially) reenters the brane depends on the global structure
of the brane worldsheet.

In this paper, we articulate the global structure of the
five-dimensional bulk for the cosmological evolution of DGP
braneworlds as a first step in understanding how large scale structure
in the universe is modified.  After reviewing the particulars of the
model, we take the equations laid out in \cite{Deffayet} and show how
one may interpret the evolution of the braneworld as a
relativistically expanding bubble, viewed either from the interior or
the exterior, depending on the specific phase of the theory.  We then
go on to examine the cosmological time foliations of the bulk and show
how peculiarities arise, such as shortcuts through the bulk (leading
to effective brane causality violation\footnote{Such shortcuts seem to
be prevalent in braneworld theories.  For previous studies of this
phenomenon and how such shortcuts relate to four-dimensional Lorentz
symmetry violation, see for example
\cite{Chung:1999xg,Csaki:2000dm,Caldwell:2001ja,Deffayet:2001aw}.}),
and breakdowns of that foliation occur in different regimes of the
theory, and discuss how understanding the global geometry of DGP
braneworld evolution may offer insight into cosmological
perturbations.

\section{Preliminaries}

\subsection{The Model and Cosmology}

Consider a braneworld theory of gravity with an infinite volume bulk
and a metastable brane graviton \cite{Dvali:2000hr}.  We take a
four-dimensional braneworld embedded in a five-dimensional Minkowski
spacetime.  The bulk is empty; all energy-momentum is isolated on the
brane.  The action is
\be
S_{(5)} = -\frac{1}{2}M^3 \int d^5x
\sqrt{-g}~R +\int d^4x \sqrt{-g^{(4)}}~{\cal L}_m + S_{GH}\ .
\label{action}
\ee
The quantity $M$ is the fundamental five-dimensional Planck scale.  The
first term in Eq.~(\ref{action}) corresponds to the Einstein-Hilbert
action in five dimensions for a five-dimensional metric $g_{AB}$ (bulk
metric) with Ricci scalar $R$.  The term $S_{GH}$ is the Gibbons--Hawking
action.  In addition, we consider an intrinsic
curvature term which is generally induced by radiative corrections by
the matter density on the brane \cite{Dvali:2000hr}:
\be
-\frac{1}{2}M^2_P \int d^4x \sqrt{-g^{(4)}}\ R^{(4)}\ .
\label{action2}
\ee
Here, $M_P$ is the observed four-dimensional Planck scale (see
\cite{Dvali:2000hr,Dvali:2001gm,Dvali:2001gx} for details).
Similarly, Eq.~(\ref{action2}) is the Einstein-Hilbert action for the
induced metric $g^{(4)}_{\mu\nu}$ on the brane, $R^{(4)}$ being its
scalar curvature.  The induced metric is\footnote{
	Throughout this paper, we use $A,B,\dots = \{0,1,2,3,5\}$ as
	bulk indices, $\mu,\nu,\dots = \{0,1,2,3\}$ as brane spacetime
	indices, and $i,j,\dots = \{1,2,3\}$ as brane spatial indices.}
\be
g^{(4)}_{\mu\nu} = \partial_\mu X^A \partial_\nu X^B g_{AB}\ ,
\label{induced}
\ee
where $X^A(x^\mu)$ represents the coordinates of an event on the brane
labeled by $x^\mu$.

The general time-dependent line element under consideration is of the form
\begin{equation} \label{cosmback}
ds^{2} = -N^{2}(t,y) dt^{2}
         +A^{2}(t,y)\gamma_{ij}dx^{i}dx^{j}
         +B^{2}(t,y)dy^{2}\ ,
\label{metric}
\end{equation}
and, the metric components are given by \cite{Deffayet}
\begin{eqnarray}
N &=& 1 + \epsilon|y| \ddot{a}\left(\dot{a}^2+k \right)^{-1/2} \nonumber \\
A &=& a + \epsilon|y| \left(\dot{a}^2+k \right)^{1/2}
\label{bulkmet} \\
B  &=& 1\ , \nonumber
\end{eqnarray}
where $k = -1,0$ or $1$ is the intrinsic spatial curvature parameter.
We take the total energy-momentum tensor which includes matter
and the cosmological constant on the brane to be
\begin{equation}
T^A_B|_{\rm brane}= ~\delta (y)\ {\rm diag}
\left(-\rho,p,p,p,0 \right)\ .
\end{equation}
When the matter content on the brane is specified, the induced scale
factor $a \equiv A(t,y=0)$ is determined by the Friedmann
equations \cite{Deffayet}:
\be
\sqrt{H^2+\frac{k}{a^2}} = {1\over 2r_0}
	\left[\ \epsilon + \sqrt{\frac{4r_0^2}{3M_P^2}\rho + 1}\ \right]
\label{friedmann1}
\ee
and
\be
\dot{\rho} + 3(\rho+p)H = 0\ ,
\label{friedmann2}
\ee
where we have used the usual Hubble parameter $H = {\dot a \over a}$
and we have defined a crossover scale
\be
	r_0 = {M_P^2 \over 2M^3}\ .
\label{r0}
\ee
This scale characterizes that distance over which metric fluctuations
propagating on the brane dissipate into the bulk \cite{Dvali:2000hr}.

\subsection{The Coordinate Transformation}

Starting with Eqs.~(\ref{metric}--\ref{bulkmet}) when $k=0$,
Deruelle and Dolezel \cite{Deruelle:2000ge} obtained an explicit
change of coordinate $Y^A = Y^A(X^A)$ to go to the canonical
five-dimensional Minkowskian metric
\begin{equation}
ds^2 = -(dY^0)^2 + (dY^1)^2 + (dY^2)^2 + (dY^3)^2 + (dY^5)^2.
 \label{canon}
\end{equation}
See also Ref. \cite{Deffayet}.  The coordinate transformations is
\begin{eqnarray}
Y^0 &=& A(y,t) \left( \frac{r^2}{4}+1-\frac{1}{4\dot{a}^2} \right)
-\frac{1}{2}\int dt \frac{a^2}{\dot{a}^3}~
{d\ \over dt}\left(\frac{\dot{a}}{a}\right)\ ,	\nonumber	\\
Y^5 &=& A(y,t) \left( \frac{r^2}{4}-1-\frac{1}{4\dot{a}^2} \right)
-\frac{1}{2} \int dt \frac{a^2}{\dot{a}^3}~
{d\ \over dt}\left(\frac{\dot{a}}{a}\right)\ ,  \label{changun} \\
Y^i &=&  A(y,t) x^i\ ,	\nonumber
\end{eqnarray}
where $r^2= x^i x^j \eta_{ij}$, and $\eta_{ij}$ is here a flat
Euclidean three-dimensional metric.  The transformation
Eqs.~(\ref{changun}) can be generalized to nonzero spatial curvatures.
We describe that generalization in Sec.~\ref{nonzero}.

\section{The Brane Worldsheet in a Minkowski Bulk}

We wish to focus on the early universe of cosmologies of DGP
braneworlds, and in particular on the four-dimensional big bang
evolution at early times.  For clarity, we restrict ourselves to
spatially flat braneworlds ($k = 0$) and radiation domination (i.e.,
$p = {1\over 3}\rho$) such that, using
Eqs.~(\ref{friedmann1}--\ref{friedmann2}), $a(t) = t^{1/2}$ when $H
\gg r_0^{-1}$ using appropriately normalized time units.  A more
general equation of state does not alter the qualitative picture;
changes resulting from altering the spatial curvature are discussed in
Sec.~\ref{nonzero}.  Equation~(\ref{friedmann1}) shows that the early
cosmological evolution on the brane is independent of the choice of
the $\epsilon$--parameter.  However, we see that late time evolution
depends quite sensitively to the choice of $\epsilon$.  When $\epsilon
= -1$, as $H(t)$ approaches the value $r_0^{-1}$ at late times, the
evolution of the scale factor transitions between four-dimensional
Friedmann--Lema\^{\i}tre--Robertson--Walker (FLRW) behavior to
five-dimensional FLRW behavior.  We refer to this phase as the FLRW
phase.  When $\epsilon = 1$, as $H(t)$ approaches the value
$r_0^{-1}$ at late times, the asymptotic state is an empty universe
undergoing deSitter expansion with $H = r_0^{-1}$.  We refer to this
phase as the self-accelerating phase.  See Ref.~\cite{Deffayet} for
details.

The global configuration of the brane worldsheet is determined by
setting $y = 0$ in the coordinate transformation Eq.~(\ref{changun}).
We get
\bea
	Y^0 &=& t^{1/2}\left(\frac{r^2}{4}+1-t\right)
					- {4\over 3}t^{3/2} \nonumber	\\
	Y^5 &=& t^{1/2}\left(\frac{r^2}{4}-1-t\right)
					- {4\over 3}t^{3/2}
\label{Ybrane}		\\
	Y^i &=&  t^{1/2} x^i\ .		\nonumber
\eea
The locus of points defined by these equations, for all $(t,x^i)$,
satisfies the relationship
\be
	Y_+ = {1\over 4Y_-}\sum_{i=1}^3(Y^i)^2 + {1\over 3}Y_-^3\ ,
\label{branesurf}
\ee
where we have defined $Y_\pm = {1\over 2}(Y^0\pm Y^5)$.  Note that if
one keeps only the first term, the surface defined by Eq.~(\ref{branesurf})
would simply be the light cone emerging from the origin at $Y^A = 0$.
However, the second term ensures that the brane worldsheet is
timelike except along the $Y_+$--axis.  Moreover, from
Eqs.~(\ref{Ybrane}), we see that
\be
	Y_- = t^{1/2}\ ,
\ee
implying that $Y_-$ acts as an effective cosmological time coordinate
on the brane.  The $Y_+$--axis is a singular locus corresponding to
$t=0$, or the big bang.\footnote{
The big bang singularity when $r < \infty$ is just the origin $Y_- =
Y_+ = Y^i = 0$ and is strictly pointlike.  The rest of the big bang
singularity (i.e., when $Y_+ > 0$) corresponds to the pathological
case when $r = \infty$.
}

\begin{figure} \begin{center}\PSbox{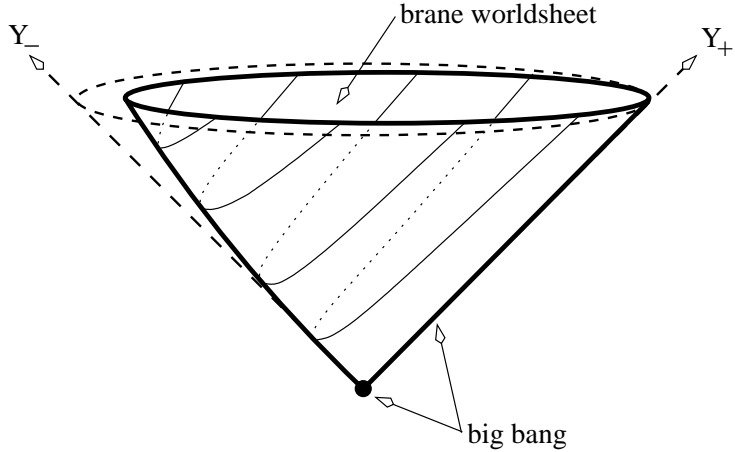
hscale=100 vscale=100 hoffset=-50 voffset=0}{3in}{2.2in}\end{center}
\caption{
A schematic representation of the brane worldsheet from an inertial
bulk reference frame.  The axes $Y_\pm$ are defined such that $Y_\pm =
{1\over 2}(Y^0\pm Y^5)$.  The bulk time coordinate, $Y^0$, is the
vertical direction, while the axis perpendicular to the
$(Y_+,Y_-)$--plane represents all $Y^i$.  The big bang is located
along the $Y_+$--axis, while the dotted surface is the future
lightcone of the event located at $Y^A = 0$.  The curves on the brane
worldsheet are examples of equal cosmological time, $t$, curves and
each is in a plane of constant $Y_-$.
}
\label{fig:brane}
\end{figure}

This picture is summarized in Fig.~\ref{fig:brane}.  Taking $Y^0$ as
its time coordinate, a bulk observer perceives the braneworld as a
compact, hyperspherical surface expanding relativistically from an
initial big bang singularity ($Y^A = 0$, for all $A$).  A single point
on that hyperspherical surface moves strictly at the speed of light
and maintains an infinite energy density.  The worldline of this
remnant of the big bang follows the $Y_+$--axis.  Note that a bulk
observer views the braneworld as spatially compact, even while a
cosmological brane observer does not.  Simultaneously, a bulk observer
sees a spatially varying energy density on the brane, whereas a
brane observer sees each time slice as spatially homogeneous.

The results of this section are analogous to those found for the
global and causal structure of Randall--Sundrum II (RS2) braneworld
cosmologies \cite{Ishihara:2000nf,Ishihara:2001qe}.  However, in the
RS2 case, one only treats that part of the bulk interior to the brane
worldsheet, and the bulk is anti-deSitter.  The bulk is strictly
Minkowski for DGP braneworlds, with the associated intuitive benefits.
In particular, the five-dimensional lightcone is respected, as well as
the causal structure of the asymptotic infinities of the spacetime.
Inertial bulk observers always have trivial worldlines.

Though the brane cosmological evolution between the FLRW phase and the
self-accelerating phase is indistinguishable at early times, the bulk
metric Eqs.~(\ref{metric}--\ref{bulkmet}) for each phase is quite
distinct.  That distinction has a clear geometric interpretation: The
FLRW phase ($\epsilon = -1$) corresponds to that part of the bulk
interior to the brane worldsheet, whereas the self-accelerating phase
($\epsilon = 1$) corresponds to bulk exterior to the brane
worldsheet.  The full bulk space is two copies of either the interior
to the brane worldsheet (the first phase) or the exterior (the latter
phase), as imposed by ${\cal Z}_2$--symmetry.  Those two copies are
then spliced across the brane.  We examine each case separately.

\section{Global Structure of the FLRW Phase}\label{FLRWphase}

\subsection{Foliation of Equal Cosmological Time Surfaces}

\begin{figure} \begin{center}\PSbox{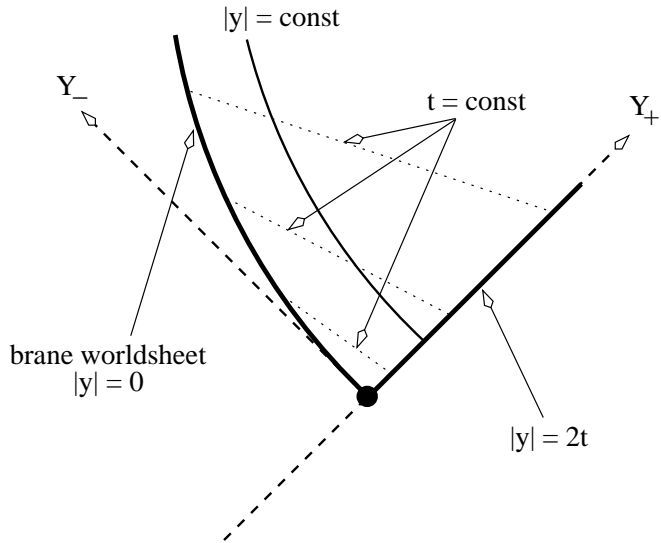
hscale=100 vscale=100 hoffset=-20 voffset=0}{3in}{2.7in}\end{center}
\caption{
A schematic representation of the FLRW brane worldsheet from an
inertial bulk reference frame with all $Y^i$--coordinates suppressed.
Equal--$|y|$ curves (thin solid lines) are roughly parallel to the
brane worldsheet (thick solid line).  The dotted lines represent
curves of equal cosmological time (equal--$t$).  Note that equal--$t$
curves intersect the big bang singularity along the $Y_+$--axis when
$|y| = 2t$, and thus extend no further in principle.  I.e., the bulk
is limited to $|y| \le 2t$, implying an observer off the braneworld
sees a finite-volume bulk.
}
\label{fig:int}
\end{figure}

If one takes $\epsilon = -1$ in Eq.~(\ref{bulkmet}), the scale
factor in a radiation dominated universe ($p = {1\over 3}\rho$)
evolves like (using appropriately normalized time units)
\bea
	a(t) = t^{1/2}\ \ \ \ \ \ \ \ \ &{\rm when}&\ \ \ \ 
						H \gg r_0^{-1}	\\
	a(t) = (r_0t)^{1/4}\ \ \ \ &{\rm when}&\ \ \ \ H \ll r_0^{-1}\ .
\eea
By taking $\epsilon = -1$, Eqs.~(\ref{changun}) implies that one is
effectively choosing the interior region of Fig.~\ref{fig:brane} as
the bulk.  Again, an observer goes from one copy of the interior to
another as one crosses the brane.  The bulk coordinates of an event
may be read from Eqs.~(\ref{changun}).  With some algebra, we arrive
at
\bea
	Y_- &=& t^{1/2}\left(1-{|y|\over 2t}\right)	\nonumber	\\
	Y_+ &=& t^{1/2}\left[\left(1-{|y|\over 2t}\right)
			\left(\frac{r^2}{4}-t\right)
					- {4\over 3}t\right]		\\
	Y^i &=& t^{1/2}x^i\left(1-{|y|\over 2t}\right)\ .	\nonumber
\eea
Figure~\ref{fig:int} depicts
equal--$|y|$ curves and equal cosmological time (equal--$t$) curves
from the point of view of an inertial bulk observer.  One can imagine
this image as a slice of Fig.~\ref{fig:brane} down the middle along
the plane determined by $Y_+$ and $Y_-$.  The bulk region that
corresponds to this phase is the region interior to the brane
worldsheet in Fig.~\ref{fig:brane}.  We see then that the volume of
any spacelike surface in the bulk is finite.  Each equal--$t$ curve
only extends to a value of $|y|=2t$.  There it intersects the big bang
singularity located along the $Y_+$--axis.

\subsection{Apparent Brane Causality Violation}

Consider an event, ${\cal O}$, located on the brane worldsheet.  We
wish to examine its past lightcone to see all events which are in
causal contact with ${\cal O}$.  Take the cosmological coordinates of
this event to be $t=t_0, y=0, x^i=0$.  If one is restricted to motion
on the brane, then geodesics are determined by the induced metric
Eqs.~(\ref{metric}--\ref{bulkmet}) with $|y| = 0$.  The locus of
past directed brane null rays are defined by the coordinates $x^i(t)$
such that
\be
	r(t) = 2(t_0^{1/2} - t^{1/2})\ ,
\ee
for a given cosmological time $t < t_0$.  As expected, even as
$t\rightarrow 0$, the past directed {\em brane} lightcone only
encompasses a finite region in $r$.  However, imagine that certain
signals (e.g., graviton degrees of freedom) may travel through the
bulk.  Since the bulk lightcone is trivial in the coordinates
$\{Y^A\}$, one can use Eqs.~(\ref{Ybrane}) to identify what events on
the brane worldsheet intersect the past {\em bulk} lightcone of ${\cal
O}$.  That locus of brane coordinates is given by the coordinates
$x^i(t)$ such that
\be
	r(t) = \sqrt{{4\over 3}{(t_0^{1/2} - t^{1/2})(t_0^{3/2} - t^{3/2})
		\over t^{1/2}t_0^{1/2}}}\ ,
\ee
where again $t$ is a cosmological time such that $t < t_0$.  Note that
as $t\rightarrow 0$, the past bulk light cone encompasses an
arbitrarily large region in comoving coordinate, $r$.  Because of the
convexity of the brane worldsheet, one can see that only the FLRW
phase allows propagation of light signals through the bulk between
events on the brane.  If signals may travel unimpeded through the
bulk, such signals would appear to travel faster than the speed of
light on the brane, thus providing an effective brane causality
violation.  One can see that this conclusion is natural by examining
Fig.~\ref{fig:brane}.  The past lightcone of any event ${\cal O}$ on
the brane worldsheet clearly encompasses the entire big bang for
$r<\infty$, which is located at the origin, $Y^A = 0$.

\section{Global Structure of the Self-Accelerating Phase}\label{selfacc}

\subsection{Foliation of Equal Cosmological Time Surfaces}

The self-accelerating phase corresponds to taking $\epsilon = 1$ in
Eq.~(\ref{bulkmet}).  With this choice of $\epsilon$,
Eq.~(\ref{bulkmet}) implies the scale factor in a radiation dominated
universe evolves as (using appropriately normalized time units)
\bea
	a(t) = t^{1/2} \ \ \ \ \ &{\rm when}&\ \ \ \ 
						H \gg r_0^{-1}	\\
	a(t) \sim e^{t/r_0}\ \ \ \ &{\rm when}&\ \ \ \ H \ll r_0^{-1}\ .
\eea
The bulk region is the region exterior to the brane worldsheet in
Fig.~\ref{fig:brane}.  The coordinates of an event in the bulk may
again be read from Eqs.~(\ref{changun}):
\bea
	Y_- &=& t^{1/2}\left(1+{|y|\over 2t}\right)	\nonumber	\\
	Y_+ &=& t^{1/2}\left[\left(1+{|y|\over 2t}\right)
			\left(\frac{r^2}{4}-t\right)
					- {4\over 3}t\right]		\\
	Y^i &=& t^{1/2}x^i\left(1+{|y|\over 2t}\right)\ .	\nonumber
\eea
Unlike the FLRW phase, spatial slices of the Minkowski bulk are
infinite in volume.  However, there exists a subtlety with equal-time
foliation of the big bang for the self-accelerating phase.  The
cosmological coordinate system Eqs.~(\ref{metric}) develops
singularities when $|y|=2t$.  This condition in the
FLRW phase defines a set of points lying on the brane itself and
corresponded to the big bang singularity.  In the self-accelerating
phase, the locus $|y|=2t$ exists in the Minkowski bulk.  Here,
the surface $|y|=2t$ follows the relationship
\be
	Y_+ = {1\over 4Y_-}\sum_{i=1}^3(Y^i)^2 - {1\over 12}Y_-^3\ .
\label{self-sing}
\ee
When this relationship is satisfied, $N=0$.  Moreover, one can show
that no value of $(y,t)$ covers a region where $Y_+$ is smaller than
the value defined by Eq.~(\ref{self-sing}).

\subsection{Reversal of Cosmological Time and $N\le 0$}

\begin{figure} \begin{center}\PSbox{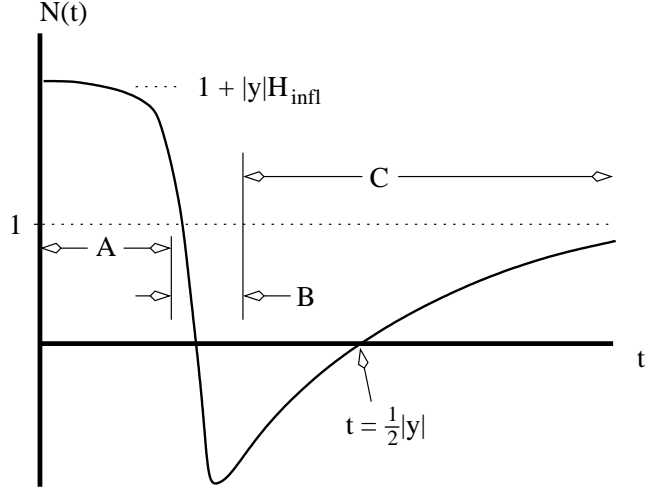
hscale=100 vscale=100 hoffset=-20 voffset=0}{3in}{2.4in}\end{center}
\caption{
The evolution of the metric function $N = 1 + |y|\ddot{a}/\dot{a}$
for a given $|y|$.  The evolution occurs in three different
qualitative phases:  (A) inflation, where the scale factor
accelerates;  (B) reheating, where the scale factor transitions
from an accelerating phase and a decelerating phase and therefore
$N$ vanishes for sufficiently large $|y|$; and (C) FLRW evolution,
where the scale factor decelerates in its growth.  In this last
phase, $N$ vanishes when $t = {1\over 2}|y|$ in a radiation
dominated universe.}
\label{fig:n}
\end{figure}

Indeed, one can show that this coordinate singularity is associated
quite generally with the existence of a decelerating scale factor and
one may derive intuition based on phenomenologically interesting
cosmic evolutions.  Consider a general scale factor evolution $a(t)$.
Then, one can ascertain the parametric evolution of equal--$|y|$ (and
fixed $r^2$) curves with respect to cosmological time, $t$.  Using the
coordinate transformation Eqs.~(\ref{changun}), one can track
$\dot{Y_+}$ and $\dot{Y_-}$ for fixed $|y|$.  We find that
\bea
	\dot{Y_+} &=& \left({r^2\dot{a}\over 4} + {1\over 4\dot{a}}\right)
			\left(1+|y|{\ddot{a}\over\dot{a}}\right)
				= \left({r^2\dot{a}\over 4}
				+ {1\over 4\dot{a}}\right)N(t,y)	\\
	\dot{Y_-} &=& \dot{a}\left(1+|y|{\ddot{a}\over\dot{a}}\right)
				= \dot{a}\ N(t,y)\ .
\eea
Thus, $\dot{Y_+}$ and $\dot{Y_-}$ vanish simultaneously when
$N(t,y)=0$, implying cusps in the equal--$|y|$ curves in the
$(Y^5,Y^0)$--plane.

Using a typical cosmological scenario, one can generate a qualitative
picture of the evolution of the metric function $N(t,y)$.  At early
times, we assume an inflationary phase that is approximately deSitter
with a Hubble scale, $H_{\rm infl}$.  In this case, the scale factor
grows with positive acceleration.  Eventually, the inflationary stage
ends and reheating occurs.  A hot FLRW-like, big bang universe is
restored.  Figure~\ref{fig:n} shows how, for a sufficiently large
$|y|$, the metric function $N(t,y)$ develops nodes.  One node occurs
during the reheating stage, the other during the FLRW stage.  When the
FLRW stage is radiation dominated, the second node occurs near $t =
{1\over 2}|y|$.  Thus, the minimum $|y|$ for which these nodes develop
depends on the earliest time at which FLRW evolution is restored.

So when $N<0$, both $Y_+$ and $Y_-$ decrease with increasing
cosmological time, $t$.  This implies a reversal of the direction of
future-versus-past for the bulk observer relative to a brane observer.
This is not surprising since one can think of $N$ acting as a sort of
speed of light.  When that value is negative, one is effectively
reversing the sign of time.

\begin{figure} \begin{center}\PSbox{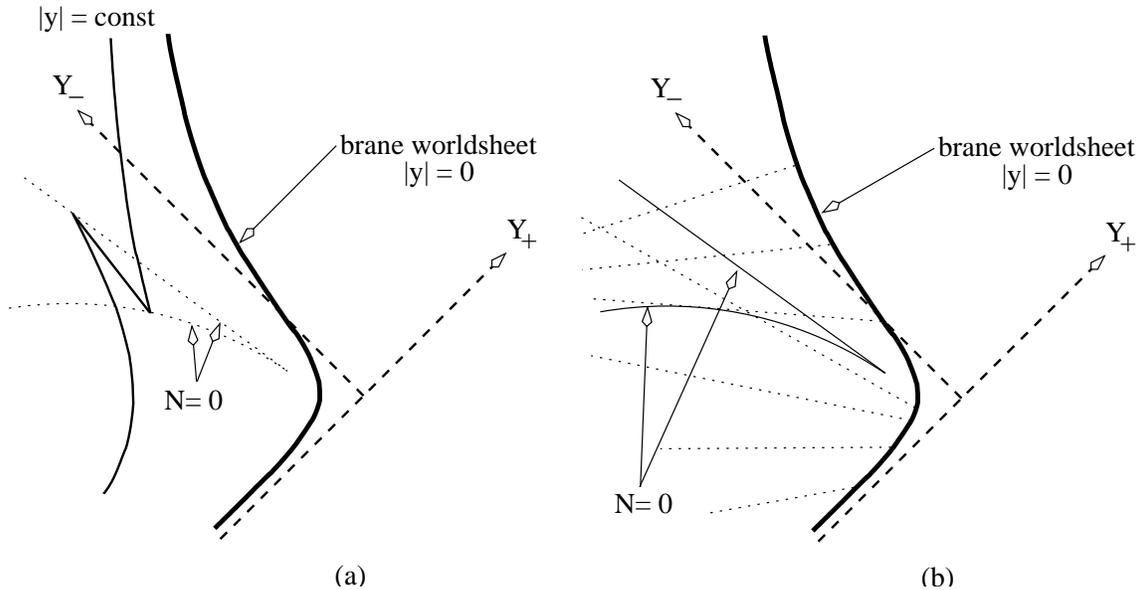
hscale=100 vscale=100 hoffset=20 voffset=0}{7in}{2.9in}\end{center}
\caption{
A schematic representation of the self-accelerating brane worldsheet
from an inertial bulk reference frame with all $Y^i$--coordinates
suppressed.  (a) Equal--$|y|$ curves passing through the wedge defined
by the dotted curve have two cusps.  For the part of the curve between
the two cusps, $N<0$, i.e. increasing $t$ implies evolution that is
{\em past} directed for a bulk inertial observer.  (b) Equal
cosmological time curves (dotted curves) in the same universe.  Each
spacetime point within the wedge (thin solid curve) defined by $N=0$
has three equal time curves passing through it.  Outside the wedge,
one can assign a unique cosmological time to each spacetime point.
}
\label{fig:ext}
\end{figure}

In order to provide a more intuitive description of the early
universe, let us imagine that the big bang is smoothed out by a stage
of early inflation that extends arbitrarily far into the past.
Figure~\ref{fig:ext} provides a visual description of equal--$|y|$ and
equal cosmological time curves from the point of view of an inertial
bulk observer.  Again, imagine taking a slice of Fig.~\ref{fig:brane}
down the middle, but where the brane worldsheet is smoothed out at the
origin to include an early inflating stage.  Indeed, in order to
fill out the region of bulk spacetime not covered by values of $Y_+$
less than that determined by Eq.~(\ref{self-sing}), one needs to
includes something like this early inflationary stage.

\subsection{Leakage and Depletion of Anisotropic Power}

In Sec.~\ref{FLRWphase} we saw how null worldlines through the bulk in
the FLRW phase could connect different events on the brane.  This
observation was a consequence of the convexity of the brane worldsheet
and the choice of bulk.  Conversely, if one chooses the bulk
corresponding to the self-accelerating phase, one may conclude that no
null lightray through the bulk connects two different events on the
brane.  Let us elaborate on the consequences of such an observation.

Consider again Fig.~\ref{fig:ext}.  Because the brane worldsurface
must be timelike and spatially compact, each external past-directed
and future-directed lightrays can only intersect the brane worldsheet
once.  But perturbations through the bulk must follows these
lightrays, since the action Eq.~(\ref{action}) dictates that the empty
bulk is pure Einstein gravity.  From the propagator
\cite{Dvali:2000hr,Dvali:2001gm,Dvali:2001gx,Deffayet:2002fn}, one can
see that gravitational modes on the brane are influenced only by bulk
modes at length scales of order ${\cal O}(r_0)$ or larger.  If one
imagines that the initial state of perturbations is localized around
the brane, then no bulk perturbations from the brane in the past can
reintersect the brane in the sufficiently late future, and that one
expects only a depletion of amplitude from the power spectrum of
perturbations at large scales.

This argument is subject to the validity of using linearized
perturbations on cosmological backgrounds.  But recall at the times of
interest (when $H \sim r_0^{-1}$), perturbations have condensed into
compact objects like galaxies, clusters, etc.  Study of compact
objects in the DGP braneworld theory
\cite{Lue:2001gc,Gruzinov:2001hp,Porrati:2002cp} suggest that
nonlinear effects may not necessarily decouple, even at large scales.
A more detailed study of how perturbation on differing scales interact
as well as the importance of nonlinearities in the DGP model is
necessary to draw a definitive conclusion on the matter.

\section{Nonflat Spatial Geometries}\label{nonzero}

For the metric (\ref{bulkmet}) with $k \neq 0$, one can find as a
change of coordinate $Y^A = Y^A(X^A)$ leading to the canonical
Minkowski metric (\ref{canon}).  It can be defined for $k= 1$ by
\cite{Deffayet}
\begin{eqnarray}
Y^0 &=& A(y,t)\ \tilde{y} + \tilde{z}\ , \nonumber \\
Y^5 &=& A(y,t)\ \tilde{Y}^5\ ,   \label{changdeux}  \\
Y^i &=& A(y,t)\ \tilde{Y}^i\ ,  \nonumber
\end{eqnarray}
where $\tilde{Y}^i$ and $\tilde{Y}^5$ are functions of the $x^i$ only
and verify
\be
(\tilde{Y}^5)^2 + \sum_{i=1}^3 (\tilde{Y}^i)^2 = 1,
\ee
which defines a three-dimensional $k=1$ maximally symmetric space.
For $k=-1$, we find \cite{Deffayet}
\begin{eqnarray}
Y^0 &=& A(y,t)\ \tilde{Y}^0\ , \nonumber \\ 
Y^5 &=& A(y,t)\ \tilde{y} + \tilde{z}\ , \label{changtrois} \\
Y^i &=& A(y,t)\ \tilde{Y}^i\ , \nonumber
\end{eqnarray}
where $\tilde{Y}^i$ and $\tilde{Y}^0$ are function of the $x^i$ only
and verify
\be
(\tilde{Y}^0)^2 - \sum_{i=1}^3 (\tilde{Y}^i)^2 = 1\ ,
\ee
which defines a three-dimensional $k=-1$ maximally symmetric space.
The quantities $\tilde{y}$ and $\tilde{z}$ are given by
\begin{eqnarray} \label{tildy}
\tilde{y} &=& \frac{\dot{a}}{\sqrt{\dot{a}^2 +k}}\ ,  \\
\tilde{z} &=& k \int dt\ \frac{1}{\dot{a}} {d\ \over dt} \left(
\frac{a}{\sqrt{\dot{a}^2+k}} \right)\ . \label{tildz}
\end{eqnarray}
So, we see that on the brane the coordinate $Y^0$ acts as the
cosmological time coordinate when $k=1$, whereas the coordinate
$Y^5$ acts as the cosmological time coordinate when $k=-1$.
Compare this to the case when $k=0$ and the cosmological time
is characterized by $Y_- = {1\over 2}(Y^0-Y^5)$.

The qualitative features represented in Fig.~\ref{fig:brane} are the
same except for the following.  In the $k=1$ scenario, the big bang is
a pointlike singularity at the origin, and the entire braneworld is a
hypersphere whose radius is determined by the induced scale factor
$a(t)$.  The brane worldsheet lies entirely within the light cone of
the origin ($Y^A=0$) such that the surface maintains a spherical
symmetry around the $Y^0$--axis.  Again, equal cosmological time
surfaces are loci of equal $Y^0$ on this brane worldsheet.  When
$k=-1$, the big bang is composed of two lightrays located along the
lightcone defined by $Y^5 = 0$.  Once again, the entire brane
worldsheet is located within the light cone of the origin.
Equal-cosmological-time surfaces are loci of equal $Y^5$ on the
brane worldsheet.  These pictures are directly analogous to those
found in \cite{Ishihara:2001qe}, but where again the bulk internal and
external to the brane worldsheet refer to different phases of the
theory, and each bulk is strictly Minkowski.  Moreover, the types of
subtleties shown to exist in the coordinate transformation for the
self-accelerating phase discussed in Sec.~\ref{selfacc} carry over.
In particular, the mapping of the cosmological coordinate system to
the bulk coordinate system is not one-to-one when there exist values
of $|y|$ where $N\le 0$.

\section{Concluding Remarks}

We examined the global structure of early universe cosmologies for
Dvali--Gabadadze--Porrati (DGP) braneworlds.  Two distinct phases
exist: the Friedmann--Lema\^{\i}tre--Robertson--Walker (FLRW) phase,
where late time evolution follows five-dimensional FLRW behavior, and
the self-accelerating phase, where the brane asymptotes to late time
deSitter expansion with an empty brane.  A bulk observer sees the
brane as a relativistically expanding, roughly hyperspherical bubble
emerging from a pointlike big bang.  Depending on the spatial
curvature of the {\em internal} brane cosmology, the brane is strictly
hyperspherical (positive spatial curvature, $k=1$), has one residual
big bang point on the hypersphere moving at exactly the speed of light
(flat spatial geometry, $k=0$), or two diametrically opposed residual
big bang points on the hypersphere, also moving at exactly the speed
of light (negative spatial curvature, $k=-1$).  Note that a bulk
observer perceives the brane as compact, even when a cosmological
brane observer would not (i.e., the $k=0,-1$ cases).  The bulk is two
identical copies of the space interior to this compact brane, glued
across a ${\cal Z}_2$--symmetric brane when in the FLRW phase.
Correspondingly, the bulk space is two copies of the space exterior to
the brane when in the self-accelerating phase.

The FLRW phase exhibits lightlike shortcuts through the bulk that
connect different events on the brane.  This phenomena can lead to
apparent brane causality violation and provides an opportunity for the
evasion of the horizon problem.  Unlike the big bang of conventional
four-dimensional FLRW cosmology, the past (bulk) lightcone of every
spacetime event on the brane contains the entire big bang for all
comoving coordinate radii $r < \infty$.  Phrasing this statement
another way is that the entire brane worldsheet is in the future
lightcone of the big bang for $r < \infty$, where the locus of
spacetime events of the big bang with $r < \infty$ is strictly
pointlike.  Thus, gravitons emitted from a time arbitrarily close to
the big bang may travel through the bulk and may transmit thermal
information to all parts of the brane during the evolution of the
early universe.  The difficulty in this mechanism for evading the
horizon problem is that bulk propagation of gravitons is at its least
significance precisely at those times (i.e., the early big bang) when
we wish to transmit them through the bulk.

The self-accelerating phase does not possess lightlike worldline that
connect different events on the brane.  This observation implies that
density perturbations leaving the brane via the bulk cannot reenter
the brane at some later time.  At sufficiently large scales, where
brane fluctuations are strongly coupled to bulk modes, one anticipates
that only depletion of power can occur for metric fluctuations, since
there is no mechanism for replenishment of power at these scales.
This argument implies a deficit in the power spectrum of linear density
perturbations at large scales.  The magnitude of this deficit, as well
as the validity of this analysis motivated by linearized
perturbations, are subject to more quantitative inquiry.
Nevertheless, one can see the advantage of developing an understanding
of the global structure of DGP braneworlds for gaining insight into
their phenomenological consequences.

\acknowledgements

The author would like to thank C.~Deffayet, G.~Dvali, G.~Gabadadze
and G.~Starkman for helpful discussions.  This work is sponsored in
part by NSF Award PHY-9996137 and the David and Lucille Packard
Foundation Fellowship 99-1462.

\end{document}